\documentclass[reprint,amsmath,amssymb,aps,prl,superscriptaddress, showpacs, eprint]{revtex4-2}
\usepackage[utf8]{inputenc}

\usepackage{amsmath,amssymb,ascmac,fancybox, multirow}
\usepackage{color}
\usepackage{graphicx}
\usepackage{url}
\usepackage{siunitx}
\usepackage{bm}
\usepackage{physics}
\usepackage{mathtools}
\usepackage{hyperref}
\usepackage[T1]{fontenc}
\usepackage{ulem}
\graphicspath{figures/}

\renewcommand{\vec}{\vb*}

\usepackage{makecell}

\newcommand{\chiP}{\chi}

\begin{document}
\date{\today}
\title{Universal relation between energy gap and dielectric constant}
\author{Yugo Onishi}
\affiliation{Department of Physics, Massachusetts Institute of Technology, Cambridge, MA 02139, USA}
\author{Liang Fu}
\affiliation{Department of Physics, Massachusetts Institute of Technology, Cambridge, MA 02139, USA}

\begin{abstract}
We establish a universal relation between the energy gap and the static dielectric constant for all insulating states. This relation yields an upper bound on the energy gap, which only depends on the electron density and electronic dielectric constant. We identify two types of energy gaps associated with transverse and longitudinal excitations at long wavelength, which correspond to the optical gap and the plasmon energy respectively. Their upper bounds are set by the dielectric constant and its inverse respectively.
The transverse gap bound is calculated for a wide range of materials and compared with the measured optical gap. A remarkable case is cubic boron nitride, in which the direct gap reaches \SI{72}{\percent} of the bound. 
 Our results are derived from the Kramers-Kronig relation and the $f$-sum rule, and therefore rest on general physical principles.
\end{abstract}

\maketitle

Insulating states of matter with a bulk energy gap are ubiquitous: examples include semiconductors, Mott insulators, and quantum Hall states. If the ground state is separated from excited states by a finite energy gap, zero-temperature conductivity necessarily vanishes at low frequency. In recent decades, significant progress has been made in understanding ground state properties of various types of insulating states \cite{kohn_theory_1964, Thouless1982, niu_quantized_1985, Hasan2010, resta_insulating_2011}, including quantum Hall states and topological insulators. %
On the other hand, the energy gap of insulating states has received little attention, despite its fundamental importance. The band gap of semiconductors determines their electrical and optical properties. The energy gap of topological insulators limits the temperature for observing quantized conductance.

Recently, we found a fundamental upper bound on the energy gap for (integer or fractional) Chern insulators, which is simply determined by the electron density and the ground state Chern number \cite{onishi_fundamental_2023}. 
Interestingly, this bound is saturated for quantum Hall states in Landau level systems, and is fairly tight for Chern insulators in twisted semiconductor bilayers at zero magnetic field. Thus, it offers a guiding principle for searching large-gap topological materials, and reveals a deep connection between the ground-state topology and the excitation spectrum.

In this work, by exploring the connection between ground state property, thermodynamic response and energy gap, we derive an upper bound on the energy gap of insulating states based on the dielectric constant. We identify two energy gaps in electronic solids that are associated with excitations that couple to longitudinal and transverse electric field respectively. We find two distinct sum rules for the dielectric constant, from which the bounds for longitudinal and transverse gaps are derived. Our gap bound applies to all insulators---topological or trivial. We show that the bound is practically useful for a wide range of materials, especially wide-bandgap semiconductors. A remarkable example is cubic boron nitride (c-BN), where the direct gap reaches up to \SI{72}{\percent} of our gap bound!

We start by considering generalized optical weights for insulating states \cite{onishi_fundamental_2023}, defined as the negative moments of the absorptive part of the optical conductivity $\sigma(\omega)$,  
\begin{align}
W_{\alpha \beta}^i  \equiv \int_0^{\infty} \dd{\omega} \frac{\sigma^{\rm abs}_{\alpha\beta}(\omega)}{\omega^i},  
 \label{eq:generalized}
\end{align}
where $\sigma^{\rm abs}_{\alpha \beta} \equiv (\sigma_{\alpha \beta} + \sigma^*_{\beta \alpha})/2$ is a Hermitian tensor consisting of the real part of longitudinal optical conductivity and the imaginary part of optical Hall conductivity. For example, $\sigma^{\rm abs}_{\alpha \beta} = \Re(\sigma_{xx}) \delta_{\alpha \beta} + i \Im(\sigma_{xy}) \epsilon_{\alpha \beta}$ in isotropic two-dimensional systems.

The zero-th moment $\Re [W^0] = \int_0^{\infty} \dd{\omega} \Re \sigma^{\rm abs}$ is the standard optical spectral weight. It is well-known from the $f$-sum rule that optical spectral weight is a ground state property given by~\cite{Kubo1957a} 
\begin{align}
    \Re W^0_{\alpha \beta} &= \frac{\pi}{2V} \expval{\eval{\pdv[2]{\hat{H}(\vec{A})}{A_\alpha}{A_{\beta}}}_{\vec{A}=0}}, 
    \label{eq:general_optical_weight} 
\end{align}
where $\hat{H}(\vec{A})$ is the Hamiltonian of the system coupled to a uniform vector potential $\vec{A}$ and $V$ is the volume of the system.  %
The fundamental microscopic Hamiltonian of electronic solids includes electron's kinetic energy $p^2/(2m)$, but no other terms involving $p^n$ with $n\geq 2$ (where $\vec p$ is the momentum). Then,          
Eq.~\eqref{eq:general_optical_weight} reduces to the universal $f$-sum rule 
\begin{align}
    W^0_{\alpha\alpha}= \pi ne^2/(2m), \label{eq:f-sum}
\end{align}
where $n$ is the electron density (including core electrons), $m$ is the bare mass, and $e(<0)$ is the charge of electron. 

For systems with a finite energy gap in the bulk, the negative-$i$-th moment $W^i$ is well-defined for any $i\geq 1$. 
Previously, we derived the topological gap bound by utilizing $W^1$, including both real and imaginary parts~\cite{onishi_fundamental_2023}. The real part of $W^1$ (termed ``quantum weight'') is related to the quantum metric of the ground states over twisted boundary conditions~\cite{souza_polarization_2000}, while its imaginary part is related to the quantized dc Hall conductivity, or the many-body Chern invariant.  %
We also note that the positive second moment was previously discussed in relation to optical properties and charge distribution~\cite{hopfield_sum_1970}.

In this work, we shift our focus to the negative-second moment of optical conductivity, $W^2$. Since the charge current is the time derivative of the polarization, the optical conductivity of insulating states is directly related to the polarizability as:
\begin{align}
    -i\omega \chiP(\omega) = \sigma(\omega). \label{polarizability}
\end{align}
Here $\sigma$ and $\chiP$ describe respectively the current and polarization induced by an {\it applied} uniform  electric field: ${\bm j}= \sigma \vec{E}_{0}$ and
${\bm P}= \chiP \vec{E}_{0}$, where ${\vec E}_0$ couples to the position of charges  
$- e {\vec E}_0 \cdot \sum_i {\vec r}_i$. 
Substituting this relation into Eq.~\eqref{eq:generalized}, we can rewrite $\Re W^2$ as 
\begin{align}
    \Re W^2  
    = \frac{1}{2}\int_{-\infty}^{\infty} \dd{\omega} \frac{\Im\chiP(\omega)}{\omega}.  
\label{eq:W2_chiP1}
\end{align}
The Kramers-Kronig relation for $\chi$ relates the right hand side of Eq.(\ref{eq:W2_chiP1}) to the thermodynamic response function $\chi^0$ which measures the change of polarization in response to an externally applied static electric field: 
\begin{eqnarray}
\chiP(0) = - \lim_{\omega \rightarrow 0} \frac{\Im \sigma}{\omega}  \equiv \chiP^0.      
\end{eqnarray}
This leads to a sum rule relating optical conductivity and thermodynamic polarizability:    
\begin{align}
  \int_0^{\infty} \dd{\omega} \frac{\Re \sigma(\omega)}{\omega^2} = \frac{\pi}{2}\chiP^0.  \label{eq:W2_chiP}
\end{align}

Using generalized optical weights, we can derive a series of upper bounds on the optical gap $E_g$ of insulating states. $E_g \equiv \hbar \omega_g $ is defined by the threshold frequency $\omega_g$ for optical absorption %
so that $\Re\sigma(\omega < \omega_g)=0$.
Since $\Re\sigma(\omega)$ is always semipositive, we can find an upper bound on the generalized optical weights as 
    $W^{i}_{\alpha\alpha} \le (\hbar/E_g)^i W^0_{\alpha\alpha}$, where $\alpha$ can be any direction. 
This inequality immediately leads to a bound on the optical gap $E_g$ given by $W^0$ and $W^i$:
\begin{align}
    E_g \le \hbar \qty(\frac{W^0_{\alpha\alpha}}{W^i_{\alpha\alpha}})^{1/i}. \label{eq:bound_Wi}
\end{align}
This inequality applies to any system at all dimensions, independent of any microscopic details.

We also note that the optical gap $E_g$ is, by definition, greater than or equal to the spectral gap $\Delta$, which is defined as the energy difference between the ground state and the first excited state. Therefore, our optical gap bound, Eq.~\eqref{eq:bound_Wi}, also gives an upper bound on the spectral gap. For example, in noninteracting band insulators, the optical gap is the minimum direct gap, while the spectral gap
can be indirect and thus smaller. In the presence of the Coulomb interaction, the optical gap and the spectral gap can be influenced by the excitonic effects.
Note that both the optical gap and the spectral gap are defined with charge-neutral states of the system and should not be confused with the electronic gap, which is the energy cost to add or remove one electron.

Setting $i=2$ in Eq.~\eqref{eq:bound_Wi} and using Eq.~\eqref{eq:W2_chiP}, we find a gap bound based on the polarizability $\chi^0$: 
\begin{align}
    E_g \le \hbar \sqrt{\frac{2W^0_{\alpha\alpha}}{\pi \chiP^0_{\alpha \alpha}}} \label{eq:epsilon_bound}.
\end{align}
For systems where the universal $f$-sum rule for $W^0$, Eq.~\eqref{eq:f-sum}, is applicable, the bound simplifies to  
\begin{align}
    E_g \le \hbar \sqrt{\frac{n e^2}{m \chiP^0_{\alpha \alpha}}} %
    \label{eq:universal_epsilon_bound}
\end{align}
This inequality applies to any gapped system with a conserved total particle number.

From now on, we shall focus on three-dimensional charged systems with long-range Coulomb interaction, such as electrons in real solids. Our goal is to relate the thermodynamic response $\chi^0$ to experimentally observable quantities. 
In order to proceed properly, we need to consider wavevector- and frequency-dependent conductivity $\sigma({\vec q}, \omega)$ and take the limit $\vec q \rightarrow 0$ with great care. It is also important to note that throughout this work, $\sigma({\vec q}, \omega)$ is defined by the current in response to the {\it externally applied} electric field ${\vec E}_0$, which we shall refer to as external conductivity.  
On the other hand, the physical conductivity as experimentally measured, which we denote as $\sigma_p$, is defined by the current in response to the {\it total} electric field $\vec E$. $\vec E$ includes the external field ${\vec E}_0$ and possibly additional field from induced charge density, i.e., due to the screening effect.

To proceed, we decompose the external conductivity $\sigma({\vec q}, \omega)$ into longitudinal and transverse parts: 
\begin{eqnarray}
\sigma_{\mu \nu}({\vec q}, \omega) = \frac{q_\mu q_\nu}{q^2} \sigma^L_{\mu \nu}(\vec q, \omega) + 
\qty(\delta_{\mu \nu}-\frac{q_\mu q_\nu}{q^2})\sigma^T_{\mu \nu}(\vec q, \omega), \nonumber  
\end{eqnarray}
where $\sigma^{L, T}(\vec q, \omega)$ relates the current to an applied electric field ${\vec E}_0$ parallel or perpendicular to the wavevector $\vec q$, respectively. In the limit $\vec q \rightarrow 0$, although the same physical conductivity $\sigma_p$ is obtained for longitudinal and transverse cases, one generally gets different results for $\sigma^L$ and $\sigma^T$. Since the transverse electric field does not couple to charge density directly, the transverse conductivity $\sigma^T({\vec q}\rightarrow 0, \omega)$ is equal to the physical conductivity at ${\vec q}=0$: 
$\sigma^T({\vec q}\rightarrow 0, \omega) = \sigma_p(\omega)$. However, the longitudinal conductivity $\sigma^L({\vec q}\rightarrow0, \omega)$ is generally different from $\sigma_p(\omega)$, because a longitudinal electric field induces charge density, which screens the applied electric field. Due to this screening effect, the total electric field is reduced from the external field by the complex dielectric constant $\epsilon$: ${\bm E}_0= \epsilon {\bm E}$. Therefore we have 
$\sigma^L({\vec q}\rightarrow 0, \omega) = \sigma_p(\omega)/\epsilon(\omega)$ for isotropic systems. The same argument holds for anisotropic systems with well-defined principle axes, in which $\epsilon$ takes a diagonal form for all frequencies.

Further noting the defining relation between physical conductivity and complex dielectric constant  
\begin{eqnarray}
\epsilon(\omega)  = 1+ \frac{ i \sigma_p (\omega)}{\epsilon_0 \omega},     
\end{eqnarray}  
we can relate transverse and longitudinal external conductivities to $\epsilon(\omega)$:  
$\sigma^T({\vec q}\rightarrow 0, \omega) = - i \omega \epsilon_0 (\epsilon(\omega) -1 )$, $ 
\sigma^L({\vec q}\rightarrow 0, \omega) = - i \omega \epsilon_0 (1 - \epsilon^{-1}(\omega))$.
These general relations were originally established using diagrammatic methods by Ambegaokar and Kohn~\cite{ambegaokar_electromagnetic_1960}; related discussions can be found in Refs.~\cite{kubo_fluctuation-dissipation_1966, eykholt_extension_1986}. For our purpose, it is useful to consider transverse and longitudinal polarizabilities of insulating states defined by the corresponding external conductivities as in Eq.~\eqref{polarizability}: 
\begin{eqnarray}
\chi^{T}(\omega) &\equiv& \frac{i \sigma^{T}(\vec{q}\to0, \omega)}{\omega} = \epsilon_0(\epsilon(\omega)-1), \label{chiT} \\
\chi^{L}(\omega) &\equiv& \frac{i \sigma^{L}(\vec{q}\to0, \omega)}{\omega} = \epsilon_0(1-\epsilon^{-1}(\omega)), \label{chiL} 
\end{eqnarray}
which are related to the complex dielectric constant $\epsilon(\omega)$ and its inverse $\epsilon^{-1}(\omega)$,  respectively.
As response functions of insulators, both $\chi^T$ and $\chi^L$ are analytic in the upper half plane of $\omega$ including at $\omega=0$ (where they are finite), and tend to zero as $|\omega|\rightarrow \infty$ (note $\epsilon\rightarrow 1$ at high frequency). Therefore,  
we can apply the Kramers-Kronig relation to them: 
\begin{eqnarray}
\int_{-\infty}^{\infty}\dd{\omega} \frac{\Im\chi^{L,T}(\omega)}{\omega} = \pi \Re\chi^{L,T}(0). \label{KKchi}
\end{eqnarray}

Combining Eq.~\ref{KKchi} with Eqs.~\eqref{chiT} and ~\eqref{chiL}, we obtain two 
sum rules:   
\begin{align}
\frac{1}{\epsilon_0 }\int_{-\infty}^{\infty} \dd \omega \frac{\Re \sigma^T (\vec q\rightarrow0, \omega )}{\omega^2}
&= \int_{-\infty}^{\infty} \dd \omega \frac{\Im \epsilon(\omega )}{\omega} \nonumber \\
&= \pi (\epsilon - 1)  
\label{Tsum}
\end{align}
and
\begin{align}
\frac{1}{\epsilon_0} \int_{-\infty}^{\infty} \dd \omega \frac{\Re \sigma^L (\vec q\rightarrow0, \omega )}{\omega^2} &= - \int_{-\infty}^{\infty} \dd \omega 
\frac{\Im \epsilon^{-1}(\omega )}{\omega} \nonumber \\
&= \pi(1- \epsilon^{-1}) %
\label{Lsum}
\end{align}
where $\epsilon=\epsilon(\omega=0)$ is the static dielectric constant. 
It is remarkable that these sum rules associated with longitudinal and transverse external conductivity give us sum rules for the imaginary part of $\epsilon(\omega)-1$ and $\epsilon^{-1}(\omega)-1$, respectively. We emphasize this is the consequence of the relation between complex dielectric constant and the response function to external perturbation $\chi^{L,T}$ shown in Eq.~\eqref{chiT} and ~\eqref{chiL}. The causality of the response function allows us to apply Kramers-Kronig relation.  
These sum rules for $\epsilon$ and $\epsilon^{-1}$ were previously derived   
from the mathematical property of the complex dielectric constant in Landau and Lifshitz~\cite{landau2013electrodynamics} (see also Refs.~\cite{altarelli_superconvergence_1972, smith_finite-energy_1978}). Our derivation above reveals their relations to the response to transverse and  longitudinal electric fields.

Combining the above sum rules for $W^2$ and the standard $f$ sum rule for $W^0$ allows us to derive upper bounds on two distinct energy gaps of real solids. In gapped systems, the real part of conductivity is nonzero only above a threshold frequency, thus taking the form  
\begin{eqnarray}
\Re \sigma^{L, T} ({\vec q}\rightarrow 0, \omega) = 0  \textrm{ for }  \hbar \omega < E^{L,T}_g. \label{LTgap}
\end{eqnarray}
This expression defines two energy gaps $E^{L,T}_g$ associated with the lowest energy excitations at ${\vec q} \rightarrow 0$ that couple to transverse and longitudinal electric fields, respectively.  By studying the response of a charged system to a transverse or longitudinal electric field, we can probe transverse and longitudinal excitations respectively. 

It is important to note that for three-dimensional charged systems, the transverse and longitudinal gaps are generally different because of the long-range Coulomb interaction. A well-known example is the Wigner crystal state of electrons in uniform positive charge background. This system has a longitudinal phonon at the plasma frequency but a vanishing transverse phonon frequency at a long wavelength~\cite{clark_coulomb_1958} (in the presence of a periodic lattice potential, the transverse mode shifts to finite frequency, leading to an insulating state with an energy gap).  
A transverse collective mode is manifested as a resonance in the dielectric constant $\epsilon(\omega)$, while the longitudinal plasma mode is manifested as a zero in $\epsilon(\omega)$ or a resonance in $\epsilon^{-1}(\omega)$.

Combining the sum rule Eq.~\eqref{Tsum} with the standard $f$ sum rule Eq.~\eqref{eq:f-sum} and the general form of transverse conductivity Eq.~\eqref{LTgap}, we obtain a bound on the transverse optical gap of electronic insulators:  
\begin{align}
    E^T_g \le \hbar \sqrt{\frac{n e^2}{m \epsilon_0 (\epsilon-1 )}} = \frac{\hbar\omega_p}{\sqrt{\epsilon-1}},
    \label{eq:universal_epsilon_boundT}
\end{align}
where $\omega_p=\sqrt{ne^2/(m\epsilon_0)}$ is the plasma frequency for the electron density $n$. 
Similarly, a bound on the longitudinal gap can be derived from Eq.~\eqref{Lsum}:   
\begin{align}
    E^L_g \le \frac{\hbar\omega_p}{\sqrt{1-\epsilon^{-1}}}. 
   \label{eq:universal_epsilon_boundL}
\end{align}
By definition, $E^{T,L}_g$ must exceed or equal to the spectral gap $\Delta$. Therefore, an upper bound on $E^{T,L}_g$ is also an upper bound on the spectral gap.   

Since the static dielectric constant $\epsilon$ is usually greater than $1$, 
the bound on the longitudinal gap is larger than that of the transverse gap. This is consistent with the fact that the longitudinal mode is associated with density fluctuation and therefore costs additional electrostatic energy due to the long-range Coulomb interaction. Since electromagnetic waves are transverse, the optical conductivity of solids as experimentally measured corresponds to $\sigma^T$, and the optical gap corresponds to the transverse gap $E_g^T$. %
On the other hand, the longitudinal gap can be measured by electron energy loss spectroscopy or inelastic X-ray scattering, which probe the dynamic structure factor.

\begin{figure}
    \centering
    \includegraphics[width=\columnwidth]{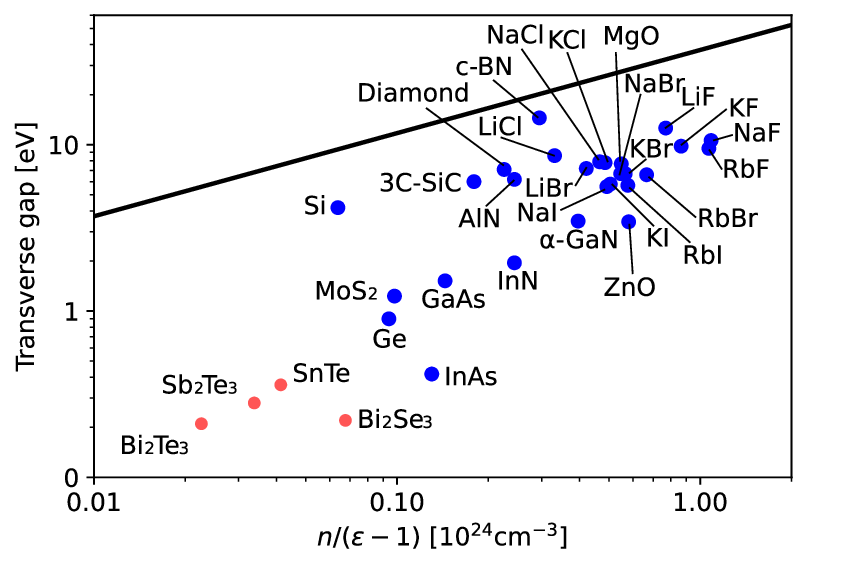}
    \caption{Universal relation between optical gap (transverse gap), dielectric constant $\epsilon$, and the electron density $n$. The black line is the transverse gap bound given by Eq.~\eqref{eq:universal_epsilon_boundT}. The gap used here is taken from optical measurements. The data points shown in red are topological insulators. } 
    \label{fig:gapbound}
\end{figure}

\begin{figure}[t]
    \centering
    \includegraphics[width=0.95\columnwidth]{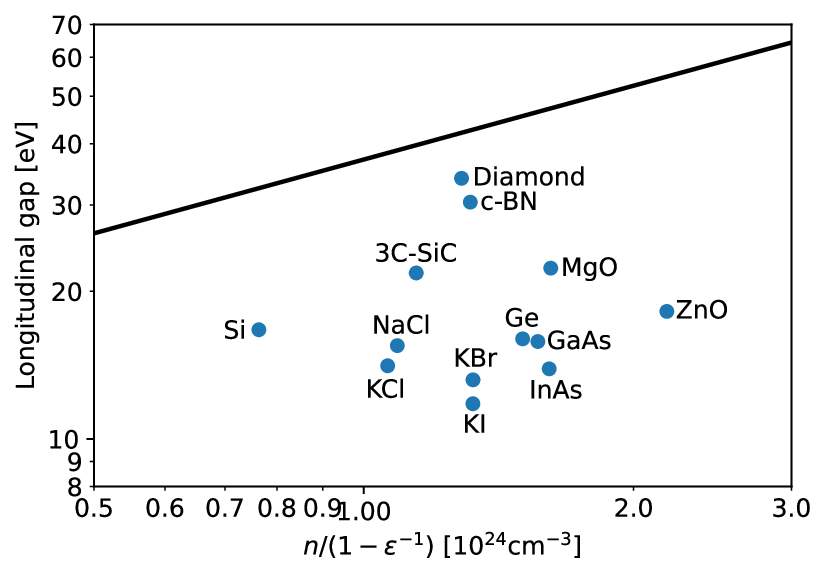}
    \caption{Universal relation between longitudinal optical gap, dielectric constant $\epsilon$, and the electron density $n$. The black line is the longitudinal gap bound given by Eq.~\eqref{eq:universal_epsilon_boundL}. The gap used here is the plasma frequency measured in electron energy loss spectroscopy.}
    \label{fig:gapbound_L}
\end{figure}

Remarkably, our gap bound given by Eqs.~\eqref{eq:universal_epsilon_boundT} and \eqref{eq:universal_epsilon_boundL} can be both practically useful. Figs.~\ref{fig:gapbound} and \ref{fig:gapbound_L} show the calculated gap bound and the experimentally measured gap for the transverse and the longitudinal gap, respectively. The two parameters here, the electron density and the dielectric constant, are given in the supplemental material. For the transverse gap, we used experimental data from optical measurements, while we used the plasmon energy measured by electron energy loss spectroscopy for the longitudinal gap. For the transverse gap bound, the most remarkable case is cubic boron nitride (c-BN) for which the gap bound given by Eq.~\eqref{eq:universal_epsilon_bound} is \SI{20.1}{\electronvolt}, while the measured direct gap is $\SI{14.5}{\electronvolt}$~\cite{madelung_semiconductors_2004}. Therefore, the actual direct gap reaches \SI{72}{\percent} of the gap bound! Besides c-BN, the measured gap in silicon, diamond, and silicon carbide (3C-SiC) reaches 45, 40, \SI{38}{\percent} of the gap bound, respectively. 
The longitudinal gap bound~\eqref{eq:universal_epsilon_boundL} can work even better than the transverse gap bound for some materials. For diamond, the measured longitudinal gap is $E_g^L = \SI{34.0}{\electronvolt}$, yielding \SI{81}{\percent} of the longitudinal gap bound \SI{42.1}{\electronvolt}. The measured longitudinal gap in c-BN, silicon, 3C-SiC reaches 72, 52, $\SI{55}{\percent}$ of the gap bound. It is remarkable that both of our gap bounds~\eqref{eq:universal_epsilon_boundT}, \eqref{eq:universal_epsilon_boundL} merely determined by electron density and dielectric constant can be fairly tight for real materials.

We now discuss various circumstances in which the gap bound~\eqref{eq:universal_epsilon_boundT} is (nearly) saturated. The first case is insulating states where electrons are strongly localized, i.e., quantum fluctuation in electron position is small.  
In the classical limit, the ground state is obtained by minimizing the sum of potential energy and electron-electron interaction energy $H_c = \sum_i V(x_i)+ \sum_{i,j} U(x_i - x_j)$ with $V$ the potential and $U$ the two-body interaction. %
When electrons are slightly displaced from their equilibrium position by $\delta x_i$, the change in the energy takes a quadratic form $H_c=E_c + \sum_{i,j}K_{ij}\delta x_i \delta x_j/2$. Diagonalizing $K$, we obtain the normal modes $x'_{\alpha}=\sum_i c_{\alpha i} \delta x_i$ (with $\sum_i c_{\alpha i} c_{\beta i}=\delta _{\alpha\beta}$) and the spring constant $K_{\alpha}$: $H_c=\sum_\alpha K_{\alpha} {x'_{\alpha}}^2/2$. %
Correspondingly, we can rewrite the total kinetic energy $\sum_i p_i^2/(2m)$ with the momentum conjugate to the normal modes, $p'_{\alpha}=\sum_i c_{\alpha i} p_i$, as $\sum_{\alpha}{p'_\alpha}^2/(2m)$. Then we obtain a collection of independent harmonic oscillators, one for each normal mode, 
\begin{align}
    H&=\sum_i \frac{{p_i}^2}{2m} + H_c \approx \sum_\alpha \frac{{p'_\alpha}^2}{2m} + \frac{1}{2}K_\alpha {x'_\alpha}^2 + \dots.
\end{align}
The frequency for each mode is $\omega_\alpha=\sqrt{K_\alpha/m}$, leading to the energy gap of $\hbar\omega_{\alpha}$ after quantization. 

To determine the polarizability of such a system, let us consider the application of a uniform static electric field $E_0$, which couples to electron displacements: $-e E_0 \sum_i \delta x_i$.  
Then, each normal mode is displaced by $\delta x'_\alpha = z_\alpha eE_0 /K_\alpha$ with $z_\alpha e\equiv\sum_i c_{\alpha i} e$ the effective charge of a mode $\alpha$, and the resulting polarization is given by $P=e\sum_i \delta x_i/V
=(e^2E_0/V)\sum_\alpha z_{\alpha}^2/K_\alpha$, hence the polarizability  is 
\begin{align}
\chi^0=\frac{e^2}{V}\sum_\alpha \frac{z_{\alpha}^2}{K_\alpha} = \frac{1}{V}\sum_\alpha \frac{e^2f_{\alpha}}{m\omega_\alpha^2},    
\end{align}
with $f_{\alpha}\equiv z_{\alpha}^2$ the oscillator strength.

Since $\sum_\alpha f_{\alpha} = N$ with the number of electrons $N$ and $K_\alpha>0$, we have $\chi^0 \le ne^2/(m\omega_0^2)$, where $\omega_0$ is the lowest oscillator frequency among all normal modes that couple to the electric field, i.e., those satisfying $f_\alpha\neq 0$. 
On the other hand, the optical gap is $E_g=\hbar \omega_0$. 
Therefore, we obtain an inequality relation between the optical gap and polarizability: $E_g \le \hbar\sqrt{ne^2/(m\chi^0)}$, recovering our bound~\eqref{eq:universal_epsilon_bound}. Note that this semiclassical result applies to both crystalline and disordered systems with localized electrons.  
From this derivation, it is clear that the bound is saturated when only a single mode couples to the electric field. This is consistent with inequality~\eqref{eq:bound_Wi} relating optical weights to the energy gap. 

As an illustrative example, let us consider a system of interacting electrons in three dimensions, which forms Wigner crystal commensurate with a lattice potential. In this case, because of the lattice translational symmetry, the normal mode can be labeled by wavevector $\vec{q}$. Writing the Fourier transform of the displacement of electrons from the equilibrium position as $\vec{u}_{\vec{q}}$, the low-energy Hamiltonian is given by 
\begin{align}
    H &=  \frac{m}{2}\sum_{\vec{q}}\abs{\dot{\vec{u}}_{\vec{q}}}^2 +  \qty(\omega_0^2 \delta_{\alpha\beta} + \omega_p^2 \frac{q_{\alpha}q_{\beta}}{q^2})u_{-\vec{q}\alpha}u_{\vec{q}\beta},
\end{align}
where $\omega_0$ is an oscillator frequency associated with the lattice potential and $\omega_p^2/q^2$ represents the Coulomb interaction between the charge density $\rho_{\vec{q}}\propto i\vec{q}\vdot\vec{u}_{\vec{q}}$.  
This system has two types of normal modes at long wavelength, a transverse mode with $\vec{u}_{\vec{q}}\perp \vec{q}$, and a longitudinal mode with $\vec{u}_{\vec{q}}\parallel\vec{q}$. 
The transverse gap (or optical gap) and the longitudinal gap (or plasmon energy) at $\vec{q}\to 0$ are respectively given by 
\begin{align}
    E_g^T=\hbar\omega_0, \quad E_g^L=\hbar\sqrt{\omega_0^2 + \omega_p^2}    
\end{align}
In this case, 
we can verify both the transverse and the longitudinal gap bounds are saturated, consistent with the fact that the external electric field couples to only a single mode (see Supplemental Materials for details).

Next, we consider the opposite limit of nearly free electron systems.   In the absence of electron-electron interaction, an approximate relation between the dielectric constant and the energy gap has been discussed in the early literature of semiconductor research~\cite{penn_wave-number-dependent_1962}. 
With certain simplifications, it was shown that $\chi^0 \approx(\hbar\omega_p/E_g)^2$ to the leading order in $E_g/E_F$ with the Fermi energy $E_F$.
Interestingly, this approximate result also leads to the saturation of our inequality~\eqref{eq:universal_epsilon_bound}.

Last but not least, two-dimensional electron gas in a magnetic field provides an interesting example of topological insulating states that saturate the bound, namely (integer or fractional) quantum Hall states. 
The Hamiltonian of this system takes the form $H=\sum_i ({\bm p} - e {\bm A})^2/(2m) + \sum_{ij} U({\bm r}_i - {\bm r}_j)$, with ${\bm A}= (By, 0)$. Due to translational invariance and the quadratic form of kinetic energy, the center-of-mass motion is decoupled from the other degrees of freedom, and     
defines a harmonic oscillator with cyclotron frequency $\omega_c=eB/m$. Since the uniform electric field only couples to the center-of-mass coordinate, we conclude by the same analysis as above that this system exactly saturates the bound. This is consistent with the Kohn theorem~\cite{kohn_cyclotron_1961}, which states that the optical absorption occurs only at $\omega=\omega_c$, leading to the saturation of Eq.~\eqref{eq:bound_Wi} and hence Eq.~\eqref{eq:universal_epsilon_bound}.

Our inequality implies that highly-polarizable electronic materials must have a small energy gap.   
This makes physical sense: the dielectric constant describes the 
polarizability of the material due to the virtual excitation of electron-hole pairs across the excitation gap~\cite{girvin_modern_2019}. 
Indeed, in narrow gap semiconductors with massive Dirac dispersion, as the gap $|\Delta|$ decreases, the dielectric constant diverges as $\sim\log|\Delta|$ in three dimensions, $1/|\Delta|$ in two dimensions, and $1/\Delta^2$ in one dimension. The asymptotic behavior in each case satisfies the inequality. Importantly, our work shows that the gap associated with transverse excitations (which couple to the electromagnetic wave) must vanish when the electronic dielectric constant $\epsilon$ diverges, but the longitudinal gap can still be finite.    
Indeed, when the system approaches Coulomb gas in three dimensions and the dielectric constant diverges, the transverse gap bound approaches zero, consistent with the presence of low-lying excitations in a metal.  
However, the plasmon excitation remains gapped, saturating the longitudinal gap bound~\eqref{eq:universal_epsilon_boundL}.

In many systems with a small energy gap, such as narrow-gap topological insulators shown in Fig.~\ref{fig:gapbound}, the electric susceptibility is large and its value is mostly determined by low-energy states near the Fermi level. On the other hand, our derivation of the universal bound on the energy gap Eq.~\eqref{eq:universal_epsilon_bound} employs the full spectral weight of all electrons in the system, which includes core electrons.  
Alternatively, we can construct an effective theory involving only low-energy states, and replace the full spectral weight $W^0$ in Eq.~\eqref{eq:epsilon_bound} with the effective spectral weight $W^0_{\rm eff}$. Since the number of electrons that participate is smaller in the effective theory, we can then expect a tighter bound for narrow gap systems. 
On this point, we can use the measured plasmon energy $E_g^L$ to %
define an effective plasma frequency $\omega_p^{\rm eff}$ associated with the effective carrier density:
$E_g^L \equiv \hbar\omega_p^{\rm eff}/\sqrt{1-\epsilon^{-1}}$.
Then, replacing the plasma frequency with $\hbar\omega_p^{\rm eff}$, the bound on optical gap  becomes 
     $E_g^T \lesssim \hbar\omega_p^{\rm eff}/\sqrt{\epsilon-1} = 
     E_g^L/\sqrt{\epsilon}$.
Interestingly, this inequality between $E_g^T, E_g^L$ and $\epsilon$ is remarkably tight for many materials (see Supplemental Materials). For example, the transverse gap of KCl, silicon, c-BN reaches \SI{82}{\percent}, \SI{87}{\percent}, and (nearly) \SI{100}{\percent} of the bound by $E_g^L$ and $\epsilon$, respectively.

In our discussion, we have not included the effect of phonons. However, due to electron-phonon coupling, the phonons can significantly contribute to the static dielectric constant, $\epsilon(0)$. Instead, the dielectric constant in our theory should be understood as the electron contribution to the dielectric constant, which is often denoted as $\epsilon(\infty)$. For example, near the phase transition to a ferroelectric phase, the static dielectric constant $\epsilon(0)$ can diverge. However, this does not necessarily imply the suppression of electronic gap %
through the bound~\eqref{eq:universal_epsilon_bound}. What is suppressed is rather the energy of optical phonons, whose softening leads to ferroelectric distortion. Generalization of our theory to include the phonon effect will be forthcoming.

\begin{acknowledgements}
We thank Raffael Resta, Ivo Souza, and Richard Martin for their comments on the manuscript. 
We are grateful to an anonymous reviewer for a stimulating suggestion.  
This work was supported by the U.S. Army Research Laboratory and the U.S. Army Research Office through the Institute for Soldier Nanotechnologies, under Collaborative Agreement Number W911NF-18-2-0048. YO is grateful for the support provided by the Funai Overseas Scholarship. LF was partly supported by Simons Investigator Award from the Simons Foundation.
\end{acknowledgements}

\bibliography{references}

\clearpage

\begin{widetext}
\appendix
\section{Supplemental Materials}
\subsection{Figures and table for the gap bounds}
\begin{figure}[htbp]
    \centering
    \includegraphics[width=0.5\columnwidth]{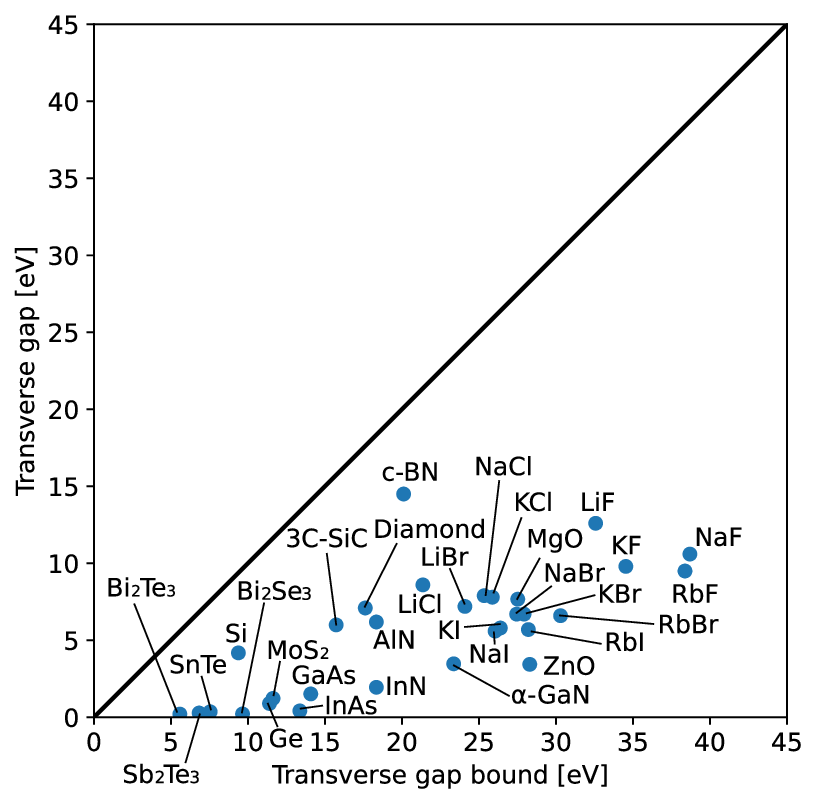}
    \caption{Measured transverse optical gap and transverse gap bound.}
    \label{fig:Tgap_vs_Tgapbound}
\end{figure}
\begin{figure}[htbp]
    \centering
    \includegraphics[width=0.5\columnwidth]{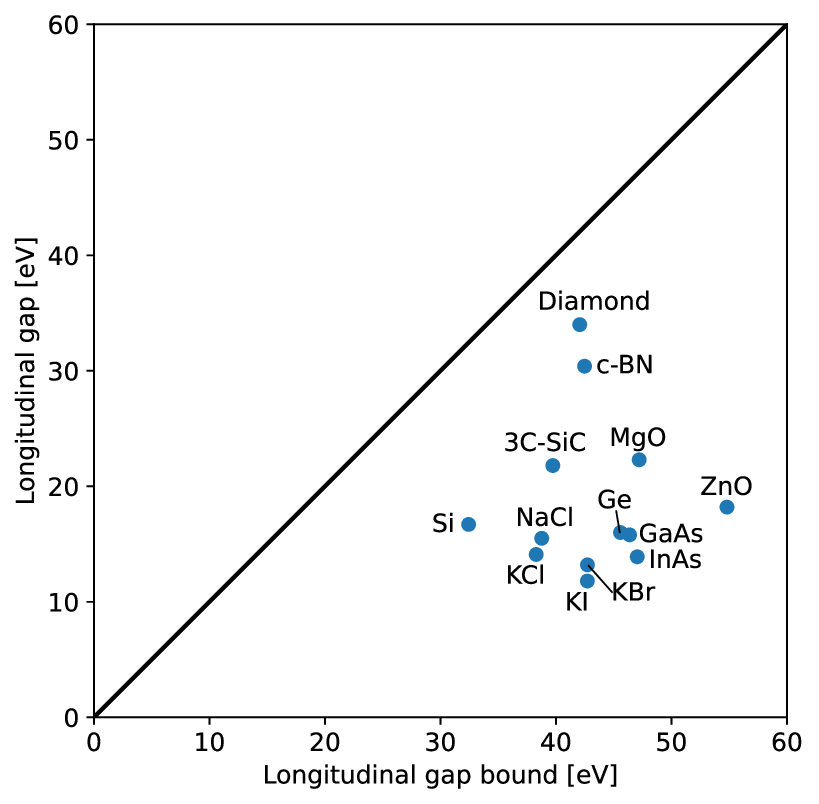}
    \caption{Measured longitudinal gap and longitudinal gap bound.}
    \label{fig:Lgap_vs_Lgapbound}
\end{figure}
\begin{figure}[htbp]
    \centering
    \includegraphics[width=0.6\columnwidth]{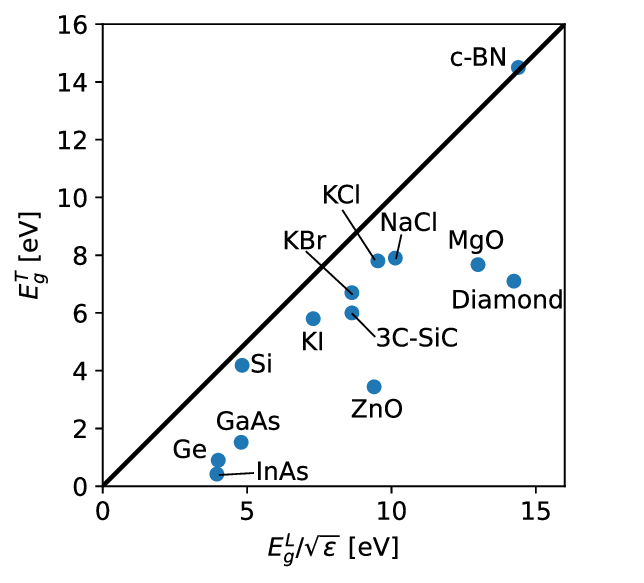}
    \caption{Approximate inequality relation between the longitudinal gap (plasmon frequency) $E_g^L$, the transverse gap (optical gap) $E_g^T$, and the dielectric constant $\epsilon$.}
    \label{fig:enter-label}
\end{figure}

\renewcommand{\arraystretch}{1.7}
\begin{table}[h!]
\centering
\begin{tabular}{l|cccccccc|l}
Material & $\epsilon(\infty)$ & $n$ [$10^{ }$\si{m^{-3}}] & $E_g^T$ [\si{\electronvolt}] & $E_g^{T,m}$ [\si{\electronvolt}] & $E_g^T/E_g^{T,m}$ & $E_g^L$ [\si{\electronvolt}] & $E_g^{L,m}$ [\si{\electronvolt}] & $E_g^L/E_g^{L,m}$ & Comment\\
\hline
c-BN & 4.46~\cite{eremets_optical_1995} & \SI{1.02}{}~\cite{madelung_semiconductors_2004}  & 14.5~\cite{madelung_semiconductors_2004}  & 20.11 & 0.721 & 30.4~\cite{jaouen_eels_1995} & 42.48 & 0.716 & $E_g^T$ is direct gap.\\
Si & 12.0~\cite{madelung_semiconductors_2004} & \SI{0.700}{}~\cite{madelung_semiconductors_2004} & 4.18~\cite{madelung_semiconductors_2004} & 9.38 & 0.446 & 16.7~\cite{egerton_electron_2008} & 32.44 & 0.515 & $E_g^T$ is direct gap.\\
Diamond & 5.7~\cite{madelung_semiconductors_2004} & \SI{1.06}{}~\cite{madelung_semiconductors_2004} & 7.1~\cite{logothetidis_origin_1992}  & 17.62 & 0.403 & 34.0~\cite{bursill_plasmon_1997} & 42.06 & 0.808 & $E_g^T$ is direct gap.\\
LiCl & 2.78~\cite{Ashcroft1976} & \SI{0.589}{}\cite{bendow_pressure_1974} & 8.6~\cite{teegarden_optical_1967} & 21.36 & 0.403 & - & 35.62 & - & $E_g^T$ is from the lowest absorption peak \\
LiF & 1.96~\cite{Ashcroft1976} & \SI{0.739}{}\cite{bendow_pressure_1974} & 12.6~\cite{roessler_electronic_1967} & 32.58 & 0.387 & - & 45.61 & - & $E_g^T$ is an excitonic gap\\
3C-SiC & 6.38~\cite{madelung_semiconductors_2004} & \SI{0.965}{}~\cite{madelung_semiconductors_2004} & 6.0~\cite{madelung_semiconductors_2004} & 15.73 & 0.381 & 21.8~\cite{costantini_analysis_2023} & 39.73 & 0.549 & $E_g^T$ is direct gap.\\
AlN & 4.93~\cite{madelung_semiconductors_2004} & \SI{0.958}{}~\cite{madelung_semiconductors_2004} & 6.19~\cite{madelung_semiconductors_2004} & 18.34 & 0.338 & - & 40.72 & - & $E_g^T$ is direct gap.. $\epsilon(\infty)=\epsilon_{\parallel}(\infty)$ \\
NaCl & 2.34~\cite{Ashcroft1976} & \SI{0.624}{}~\cite{rumble_crc_2023} & 7.9~\cite{teegarden_optical_1967} & 25.34 & 0.312 & 15.5~\cite{egerton_electron_2008} & 38.77 & 0.400 & $E_g^T$ is from the lowest absorption peak\\
KCl & 2.19~\cite{Ashcroft1976} & \SI{0.578}{}~\cite{rumble_crc_2023} & 7.8~\cite{teegarden_optical_1967} & 25.87 & 0.301 & 14.1~\cite{akkerman_inelastic_1996} & 38.29 & 0.368 & \makecell{$E_g^T$ is from the lowest absorption peak. \\ $E^L_g$ is the first prominent peak in EELS.} \\
LiBr & 3.17~\cite{Ashcroft1976} & \SI{0.914}{}\cite{bendow_pressure_1974} & 7.2~\cite{teegarden_optical_1967} & 24.09 & 0.299 & - & 42.90 & - & $E_g^T$ is from the lowest absorption peak\\
KF & 1.85~\cite{Ashcroft1976} & \SI{0.736}{}\cite{bendow_pressure_1974} & 9.8~\cite{teegarden_optical_1967} & 34.54 & 0.284 & - & 46.98 & - & $E_g^T$ is from the lowest absorption peak\\
MgO & 2.94~\cite{madelung_semiconductors_2004} & \SI{1.07}{}~\cite{madelung_semiconductors_2004} & 7.67~\cite{madelung_semiconductors_2004} & 27.52 & 0.279 & 22.3~\cite{egerton_electron_2008} & 47.21 & 0.472 & $E_g^T$ is an excitonic gap\\
NaF & 1.74~\cite{Ashcroft1976} & \SI{0.804}{}~\cite{rumble_crc_2023} & 10.6~\cite{teegarden_optical_1967} & 38.70 & 0.274 & - & 51.05 & - & $E_g^T$ is from the lowest absorption peak\\
RbF & 1.96~\cite{Ashcroft1976} & \SI{1.03}{}\cite{bendow_pressure_1974} & 9.5~\cite{teegarden_optical_1967} & 38.38 & 0.248 & - & 53.73 & - & $E_g^T$ is from the lowest absorption peak \\
NaBr & 2.59~\cite{Ashcroft1976} & \SI{0.869}{}\cite{bendow_pressure_1974} & 6.7~\cite{teegarden_optical_1967} & 27.45 & 0.244 & - & 44.18 & - & $E_g^T$ is from the lowest absorption peak\\
KBr & 2.34~\cite{Ashcroft1976} & \SI{0.758}{}\cite{bendow_pressure_1974} & 6.7~\cite{teegarden_optical_1967} & 27.93 & 0.240 & 13.2~\cite{akkerman_inelastic_1996} & 42.73 & 0.309 & \makecell{$E_g^T$ is from the lowest absorption peak. \\ $E_g^L$ is the first prominent peak in EELS.} \\
KI & 2.62~\cite{Ashcroft1976} & \SI{0.818}{}\cite{bendow_pressure_1974} & 5.8~\cite{teegarden_optical_1967} & 26.39 & 0.220 & 11.8~\cite{akkerman_inelastic_1996}& 42.72 & 0.276 & \makecell{$E_g^T$ is from the lowest absorption peak. \\ $E_g^L$ is the first prominent peak in EELS.} \\
RbBr & 2.34~\cite{Ashcroft1976} & \SI{0.892}{}\cite{bendow_pressure_1974} & 6.6~\cite{teegarden_optical_1967} & 30.30 & 0.218 & - & 46.35 & - & $E_g^T$ is from the lowest absorption peak\\
NaI & 2.93~\cite{Ashcroft1976} & \SI{0.950}{}\cite{bendow_pressure_1974} & 5.6~\cite{teegarden_optical_1967} & 26.05 & 0.215 & - & 44.58 & - & $E_g^T$ is from the lowest absorption peak\\
RbI & 2.59~\cite{Ashcroft1976} & \SI{0.918}{}\cite{bendow_pressure_1974} & 5.7~\cite{teegarden_optical_1967} & 28.21 & 0.202 & - & 45.40 & - & $E_g^T$ is from the lowest absorption peak\\
$\alpha$-GaN & 5.2~\cite{madelung_semiconductors_2004} & \SI{1.66}{}~\cite{madelung_semiconductors_2004} & 3.48~\cite{madelung_semiconductors_2004} & 23.36 & 0.149 & - & 53.27 & - &  $E_g^T$ is A-exciton. $\epsilon(\infty)=\epsilon_\perp(\infty).$\\
ZnO & 3.75~\cite{madelung_semiconductors_2004} & \SI{1.60}{}~\cite{madelung_semiconductors_2004} & 3.44~\cite{madelung_semiconductors_2004} & 28.30 & 0.122 & 18.2~\cite{huang_characterization_2011} & 54.81 & 0.332 & $\epsilon(\infty)=\epsilon_\parallel(\infty).$\\
GaAs & 10.9~\cite{madelung_semiconductors_2004} & \SI{1.42}{}~\cite{madelung_semiconductors_2004} & 1.52~\cite{madelung_semiconductors_2004} & 14.08 & 0.108 & 15.8~\cite{horak_cerenkov_2015} & 46.38 & 0.341 & $E_g^T$ is direct gap.\\
InN & 8.4~\cite{madelung_semiconductors_2004} & \SI{1.80}{}~\cite{madelung_semiconductors_2004} & 1.95~\cite{madelung_semiconductors_2004} & 18.34 & 0.106 & - & 53.15 & - & $E_g^T$ is direct gap.
\\
Ge & 16.0~\cite{madelung_semiconductors_2004} & \SI{1.41}{}~\cite{madelung_semiconductors_2004} & 0.898~\cite{madelung_semiconductors_2004} & 11.40 & 0.079 & 16.0~\cite{poursoti_deep_2022} & 45.59 & 0.351 & $E_g^T$ is direct gap.\\
SnTe & 40 \cite{suzuki_optical_1995,schoolar_optical_1968} & \SI{1.61}{}~\cite{madelung_semiconductors_2004} & 0.36~\cite{madelung_semiconductors_2004} & 7.55 & 0.048 & - & 47.75 & - & \makecell{$E_g^T$ is direct gap.  $\epsilon(\infty)$ varies \\ among Ref.~\cite{suzuki_optical_1995, schoolar_optical_1968}  around 40-50.} \\
Sb$_2$Te$_3$ & 51.0~\cite{madelung_semiconductors_2004} & \SI{1.69}{}~\cite{madelung_semiconductors_2004} & 0.28~\cite{madelung_semiconductors_2004} & 6.83 & 0.041 & - & 48.75 & - & $E_g^T$ is direct gap. $\epsilon(\infty)=\epsilon_\perp(\infty).$\\
Bi$_2$Te$_3$ & 85.0~\cite{madelung_semiconductors_2004} & \SI{1.90}{}~\cite{madelung_semiconductors_2004} & 0.21~\cite{nemov_band_2019} & 5.58 & 0.038 & - & 51.49 & - & $\epsilon(\infty)=\epsilon_\perp(\infty).$\\
InAs & 12.4~\cite{madelung_semiconductors_2004} & \SI{1.48}{}~\cite{madelung_semiconductors_2004} & 0.418~\cite{madelung_semiconductors_2004}& 13.37 & 0.031 & 13.9~\cite{kundmann_study_1988} & 47.04 & 0.295 &   $E_g^T$ is direct gap.\\
Bi$_2$Se$_3$ & 29.0~\cite{madelung_semiconductors_2004} & \SI{1.89}{}~\cite{madelung_semiconductors_2004} & 0.22~\cite{martinez_determination_2017} & 9.65 & 0.023 & - & 51.98 & - & $\epsilon(\infty)=\epsilon_\perp(\infty).$ \\
\end{tabular}
\caption{Parameters used to calculate the gap bound in Fig.~\ref{fig:gapbound}. $\epsilon(\infty)$ is the optical dielectric constant, $n$ is the electron density, $E_g^T$ is the optical gap, $E_g^L$ is the measured plasma frequency in electron energy loss spectroscopy (EELS), and $E_g^{T/L,m}$ is the gap bound calculated from Eqs.~\eqref{eq:universal_epsilon_boundT}, \eqref{eq:universal_epsilon_boundL}.}
\end{table}
\clearpage

\subsection{Dielectric constant in Wigner crystals}
\subsubsection{Classical description of Wigner crystal}
In the main text, we discussed a scenario of strongly localized electrons to (nearly) saturate our gap bound.  Such a scenario is expected to be realized in, for example, c-BN, where the possible maximum kinetic energy of the system is estimated to be $\hbar^2 n^{2/3}/(2m)\sim\SI{3.9}{\electronvolt}$, which is much smaller than the direct gap $\SI{14.5}{\electronvolt}$.

Motivated by this scenario, we consider a low-density interacting electron gas with a uniform positive charge background in three dimensions. Then the interacting energy will dominate over the kinetic energy, and thus electrons are expected to be strongly localized.  

The energy of this system is described by 
\begin{align}
    H &= \sum_{i} \frac{m}{2}\dot{\vec{r}}_i^2 + \frac{1}{2}\sum_{i\neq j} \frac{e^2}{4\pi\epsilon_0\abs{\vec{r}_i-\vec{r}_j}}.
\end{align}
where the second term is the Coulomb interaction, $\vec{r}_i$ is the position of $i$-th electron and $e(<0)$ is the charge of an electron. When the system forms Wigner crystal, the electrons are localized at $\vec{r}_i=\vec{R}_i$ so that the interaction energy is minimized. Then we can expand the Coulomb interaction term near the minima and treat them as the harmonic oscillators. $\vec{R}_i$ forms a periodic lattice, and the corresponding reciprocal lattice vectors are denoted by $\vec{G}$. Below we consider the case where each unit cell contains one electron. The number of unit cells and the volume of the system are denoted by $N$ and $V$, respectively. 

To derive the phonon dispersion, let us consider the displacements of the electrons: 
\begin{align}
    \vec{r}_i &= \vec{R}_i + \vec{u}_i.
\end{align}
Then the Hamiltonian up to the second order in the displacements is given by
\begin{align}
    &H = E_0 + \frac{1}{4}\sum_{i\neq j} (\vec{u}_i-\vec{u}_j)_{\alpha}K_{\alpha\beta}(\vec{R}_i-\vec{R}_j) (\vec{u}_i-\vec{u}_j)_{\beta} \\
    &K_{\alpha\beta}(\vec{R}) = e^2\pdv[2]{}{r_{\alpha}}{r_{\beta}}U(\vec{r}+\vec{R})|_{\vec{r}=0},
\end{align}
where $E_0$ is the energy without any displacement.
In terms of the Fourier representation, defining $\vec{u}_i = (1/\sqrt{V})\sum_{\vec{q}}e^{i\vec{q}\vdot\vec{R}_i}u_{\vec{q}}$, we have 
\begin{align}
    H &= E_0 +  \sum_{\vec{q}} \qty(\frac{m}{2}\abs{\dot{\vec{u}}_{\vec{q}}}^2 + \frac{1}{2} u_{-\vec{q}}(K_0-K(\vec{q}))\vec{u}_{\vec{q}}) \nonumber \\
    &\equiv E_0 + \sum_{\vec{q}}H({\vec{q}}).
\end{align}
Here, $K_0$ and $K(\vec{q})$ are defined as 
\begin{align}
    &K_0 \equiv \sum_{\vec{R}\neq 0} K(\vec{R}) = \frac{e^2}{4\pi\epsilon_0}\sum_{\vec{R}\neq 0}\pdv[2]{}{r_{\alpha}}{r_{\beta}}\eval{\qty(\frac{1}{\abs{\vec{r}+\vec{R}}})}_{\vec{r}=0} \\
    &K(\vec{q}) \equiv \sum_{\vec{R}\neq 0}e^{-i\vec{q}\vdot\vec{R}}K(\vec{R}) = \frac{e^2}{4\pi\epsilon_0}\sum_{\vec{R}\neq 0}\pdv[2]{}{r_{\alpha}}{r_{\beta}}\eval{\qty(\frac{e^{-i\vec{q}\vdot\vec{R}}}{\abs{\vec{r}+\vec{R}}})}_{\vec{r}=0}
\end{align}
We note that we \textit{cannot} naively interchange the sum and the derivative in the expression for $K_0$, because the sum over $\vec{R}$ without the derivative will diverge. However, we can still rewrite them by subtracting a constant in the derivative as 
\begin{align}
     &K_0 = \frac{e^2}{4\pi\epsilon_0}\pdv[2]{}{r_{\alpha}}{r_{\beta}}\eval{\sum_{\vec{R}\neq 0}\qty(\frac{1}{\abs{\vec{r}+\vec{R}}}-\frac{1}{\abs{\vec{R}}})}_{\vec{r}=0} \label{eq:K_0} \\
    &K(\vec{q}) = \frac{e^2}{4\pi\epsilon_0}\pdv[2]{}{r_{\alpha}}{r_{\beta}}\eval{\sum_{\vec{R}\neq 0}\qty(\frac{e^{-i\vec{q}\vdot\vec{R}}}{\abs{\vec{r}+\vec{R}}})}_{\vec{r}=0} \label{eq:K}
\end{align}
In this form, the quantity under the derivative does not diverge.

To calculate $K$ and $K_0$, we first recall the Ewald summation formula~\cite{ziman1972principles}:
\begin{align}
     \sum_{\vec{R}\neq 0} \frac{e^{-i\vec{q}\vdot\vec{R}}}{\abs{\vec{r}+\vec{R}}} = \frac{4\pi N}{V}\sum_{\vec{G}}\qty(\frac{e^{-(\vec{G}+\vec{q})^2/(4\alpha^2) + i(\vec{G}+\vec{q})\vdot\vec{r}}}{\abs{\vec{G}+\vec{q}}^2}) + \sum_{\vec{R}\neq0}\frac{e^{-i\vec{q}\vdot\vec{R}}}{\abs{\vec{r}+\vec{R}}} \mathrm{erfc}(\alpha\abs{\vec{r}+\vec{R}}) - \frac{1-\mathrm{erfc}(\alpha\abs{\vec{r}})}{\abs{\vec{r}}} ,
\end{align}
where $\mathrm{erfc}(x)$ is the complementary error function defined as 
\begin{align}
    \mathrm{erfc}(x) &\equiv \frac{2}{\sqrt{\pi}}\int_x^\infty e^{-t^2}\dd{t},
\end{align}
and $\alpha(>0)$ is a parameter to be chosen so that the convergence of these sums will be fast enough. 

We note that this formula diverges when $\vec{q}=0$ due to $\vec{G}=0$ term in the first term. However, this will not be a problem when one plugs this formula into Eq.~\eqref{eq:K_0} because of the subtraction. More specifically, when $\vec{q}=0$, the term with $\vec{G}=0$ vanishes because of the subtraction. In other words, only when $\vec{q}=0$, one need to exclude $\vec{G}=0$ term from the sum. This is the only difference between $K_0$ and $K(\vec{q})$ in the $\vec{q}\to 0$ limit.

Then one can easily calculate $K_0-K(\vec{q})$ in $\vec{q}\to 0$ limit as
\begin{align}
    K_{0,\alpha\beta}-K_{\alpha\beta}(\vec{q}\to 0) &= \frac{ne^2}{\epsilon_0} \frac{q_{\alpha}q_{\beta}}{q^2}
\end{align}
with $n\equiv N/V$, and hence the Hamiltonian for $\vec{q}\to 0$ limit is given by 
\begin{align}
    H({\vec{q}} \to 0) &= \frac{m}{2}\abs{\dot{\vec{u}}_{\vec{q}}}^2 + \frac{1}{2} u_{-\vec{q}} \frac{ne^2}{\epsilon_0} \frac{q_{\alpha}q_{\beta}}{q^2}\vec{u}_{\vec{q}}
\end{align}
Therefore, the dispersion for the longitudinal and transverse phonon in $\vec{q}\to 0$ limit is given by 
\begin{align}
    \omega^L(\vec{q}\to 0) &= \omega_p \equiv \sqrt{\frac{ne^2}{m\epsilon_0}}\\
    \omega^T(\vec{q}\to 0) &= 0
\end{align}

\subsubsection{Pinned Wigner crystal}
When the Wigner crystal is pinned by a harmonic potential with frequency $\omega_0$, the Hamiltonian is 
\begin{align}
    H({\vec{q}} \to 0) &= \frac{m}{2}\qty(\abs{\dot{\vec{u}}_{\vec{q}}}^2 +  \qty(\omega_0^2+\omega_p^2 \frac{q_{\alpha}q_{\beta}}{q^2})u_{-\vec{q}\alpha}{u}_{\vec{q}\beta}),
\end{align}
and the resulting dispersion at $\vec{q}\to 0$ will be 
\begin{align}
    \omega_L(\vec{q}\to 0) &= \sqrt{\omega_0^2 + \omega_p^2}, \\
    \omega_T(\vec{q}\to 0) &= \omega_0.
\end{align}

\subsubsection{Optical response and dielectric constant}
Let us calculate the optical response for the pinned Wigner crystal within the classical treatment. We consider only $\vec{q}\to 0$ mode. Since each mode behaves as the harmonic oscillator with a frequency $\omega_{L/T}$, the response of the polarization $\vec{P}=ne\vec{u}=\epsilon_0\chi \vec{E}$ to the electric field with frequency $\omega$ is given by 
\begin{align}
    \chi^{L/T}(\omega) = \frac{\omega_p^2}{\omega_{L/T}^2-\omega^2}.
\end{align}
Then the dielectric function is given by 
\begin{align}
    \epsilon(\omega) = 1+\chi^T(\omega) = \frac{\omega_L^2 - \omega^2}{\omega_T^2 - \omega^2}
\end{align}
This is consistent with another expression for the dielectric constant, $\epsilon(\omega) = (1-\chi^L(\omega))^{-1}$. 
\end{widetext}
\end{document}